\documentstyle[aps,prl,twocolumn,psfig]{revtex}

\begin{document}
\draft

\title{Instability and wavelength selection during step flow growth of metal
surfaces vicinal to fcc(001)}
\author{M. Rusanen$^{1}$, I. T. Koponen$^{2}$, J. Heinonen$^1$ and T. Ala-Nissila$%
^{1,3}$}
\address{
$^{1}$Helsinki Institute of Physics and Laboratory of
Physics, Helsinki University of Technology,
P.O. Box 1100, FIN--02015 HUT, Espoo, Finland
$^{2}$Department of Physics, University of Helsinki,
P.O. Box 9, FIN--00014 University of Helsinki, Helsinki, Finland
$^{3}$Department of Physics, Brown University, Providence R.I.
02912--1843}
\maketitle

\begin{abstract}
We study the onset and development of ledge instabilities during growth of
vicinal metal surfaces using kinetic Monte Carlo simulations. We observe the
formation of periodic patterns at [110] close packed step edges on surfaces
vicinal to fcc(001) under realistic molecular beam epitaxy conditions. The
corresponding wavelength and its temperature dependence are studied by
monitoring the autocorrelation function for step edge position. Simulations
suggest that the ledge instability on fcc(1,1,$m$) vicinal surfaces is
controlled by the strong kink Ehrlich--Schwoebel barrier, with the
wavelength determined by dimer nucleation at the step edge. Our results are
in agreement with recent continuum theoretical predictions, and experiments
on Cu(1,1,17) vicinal surfaces.
\end{abstract}


In surface growth under Molecular Beam Epitaxy (MBE) conditions, one of the
central processes that controls the morphology of the surface is mass
transport between growing layers. In homoepitaxial growth, there is usually
an additional energy barrier controlling the interlayer transitions, known
as the Ehrlich-Schwoebel (ES) or step edge barrier \cite{Sch69}. It plays a
central role in stabilization of certain growing crystalline facets, and
eventually leads to growth of mound-like structures with a dynamically
selected slope and length scale \cite{Sie94}. The influence of an interlayer
ES barrier on growth has been extensively studied and its role on growth is
now well understood \cite{Jeo99,Pol00}.

However, recently it has been realized that in $1+1$ dimensional ledge
growth corresponding to step-flow geometry, there is an analogous phenomenon
which is due to the additional {\it kink Ehrlich-Schwoebel energy barrier}
for going around a kink site at the step edge. The corresponding kink
Ehrlich-Schwoebel effect (KESE) generated by it leads to the growth of a
regular instability at step edge with a dynamically selected wavelength \cite
{Pie99}. This is in contrast to the so-called Bales-Zangwill instability
(BZI) \cite{Bal90} which tends to destabilize the 1D ledge morphology during
growth because of terrace diffusion and step crossing, with no assumptions
about line diffusion along the ledge. Ledge instabilities were originally
found and reported experimentally on Cu(1,1,17) vicinal surfaces \cite{Scw97}
but attributed to the BZI scenario. More recent experiments on the
Cu(1,1,17) surface propose that the KESE instability may lead to
formation of regularly shaped patterns with dynamical wavelength selection 
\cite{Mar99}. Recent theoretical studies of such instabilities suggest that
KESE may indeed supersede BZI in the formation of growth patterns \cite
{Pie99,Mur99}.

The growth of instabilities and wavelength selection have been studied
recently within the framework of a continuum step model \cite{Pie99} and a
simple solid-on-solid (SOS) lattice model \cite{Mur99}. For the latter case,
it has been demonstrated by kinetic Monte Carlo (KMC) simulations that KESE
leads to the formation of wavy steps \cite{Mur99}. However, so far there is
neither detailed knowledge of the actual structure of the patterns nor their
dynamical evolution under realistic MBE conditions. The predictions of the
continuum theory are mainly given in the weak KESE limit but they are not
directly applicable in interpreting the experiments \cite{Pie99,Pol96}.
Namely, for simple metals such as Cu surfaces vicinal to fcc(001), at close
packed step edges a strong KESE is expected. This is because theoretical
estimates indicate barriers of the order of 0.5 eV for jumps around the kink
site in the close packed $\left[ 110\right]$ direction \cite{Mer97}. This is
nearly twice the barrier of 0.26 eV for jumps along the straight step edge 
\cite{Mer97}. Based on the symmetry between the interlayer ES barrier which
vanishes for $\left[ 100\right]$ step edges \cite{Kur98}, it is expected
that the kink ES barrier vanishes for these orientations, too.
Therefore, on surfaces vicinal to fcc(001)
one should see a clear difference in unstable growth between the $\left[ 100%
\right]$ steps and the $\left[110\right] $ steps, the latter case being
dominated by KESE.

In this letter we present a detailed study of the growth and morphology of
the steps on the Cu(1,1,17) surface vicinal to Cu(001). The model system
used here is based on KMC simulations of a lattice-gas model \cite{Hei99}
with energetics obtained from the effective medium theory (EMT) potential 
\cite{Mer97}. As discussed in detail in Ref. \cite{Mer97} the EMT barriers
and their relative ordering is in good agreement with available experimental
data for the Cu(001) surface. The intralayer hopping rate $\nu$ of an atom
to a vacant nearest neighbor (NN) site can be well approximated by \cite
{Mer97}
\begin{equation}
\nu =\nu _{0}\exp \left[ -\beta (E_{S}+\min (0,\Delta _{NN})E_{B}\right]
\end{equation}
\psfig{file=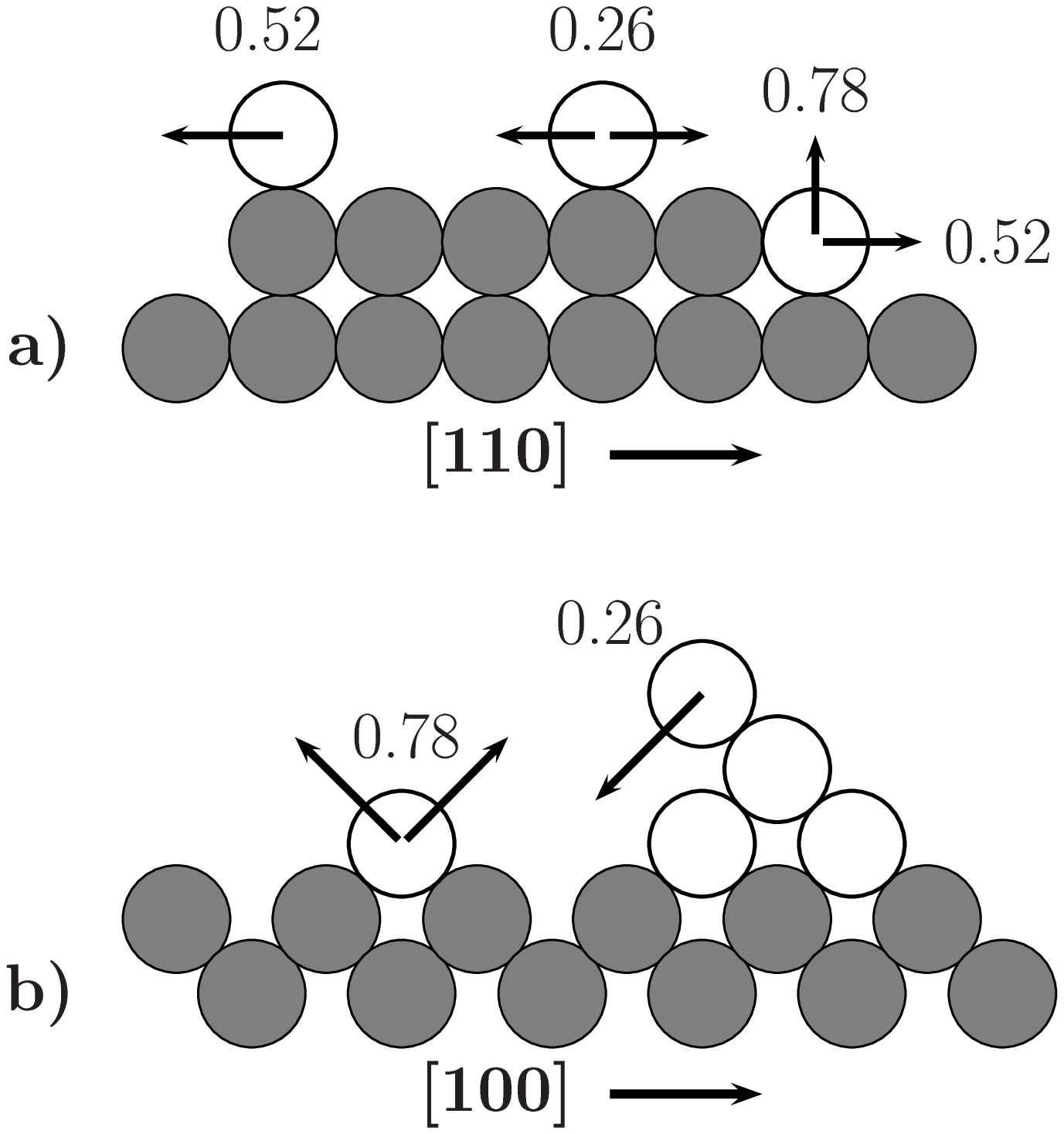,width=8.2cm}

\noindent {\bf Fig. 1}:
Ledge geometry and some relevant single atom jump processes
on (a) the close packed $\left[ 110\right]$ step edge
and on (b) the open $\left[ 100\right]$ step edge. The energy barriers
are given in eV. 

\bigskip

where the attempt frequency $\nu _{0}=3.06\times 10^{12}$ s$^{-1}$ and the
barrier for the jump of single adatom is $E_{S}=0.399$ eV. When there is at
least one atom diagonally next to the saddle point the barrier is $%
E_{S}=0.258$ eV. The change in bond number $-3 \leq \Delta _{NN}\leq 3$ is
the difference on the number of NN bonds between initial and final states
with bond energy $E_{B}=-0.260$ eV. The interlayer processes are also
included in the model. Step crossing has barrier of 0.58 eV from a straight
edge and 0.44 eV through the kink site \cite{Mer97}. We note that within the
EMT, barriers on the Ag(001) and Ni(001) surfaces are very similar to those
of Cu(001) up to an overall scaling factor \cite{Mer97}. The geometry of the
step edges with some of the pertinent single jump processes are shown in
Fig. 1. The energetics of the model give a kink barrier of $0.518$ eV for
the $\left[110\right]$ steps, and vanishingly weak KESE with a barrier of
only $0.002$ eV for the open $\left[100\right]$ steps.

The KMC simulations were implemented using the algorithm by Bortz {\it et al.%
} \cite{Bor75} with a binary tree structure \cite{Blu95}. This allowed us to
grow up to five monolayers of Cu under realistic temperature and flux
conditions to study the development of unstable ledge patterns. The explored
temperature range was $T=240 - 310$ K and the flux $F=3\times 10^{-3} - 1.0$
ML/s. Thus the ratio between the terrace diffusion and the deposition flux $%
D/F \approx 6\times 10^5 - 9\times 10^7$ in units of the lattice constant
\psfig{file=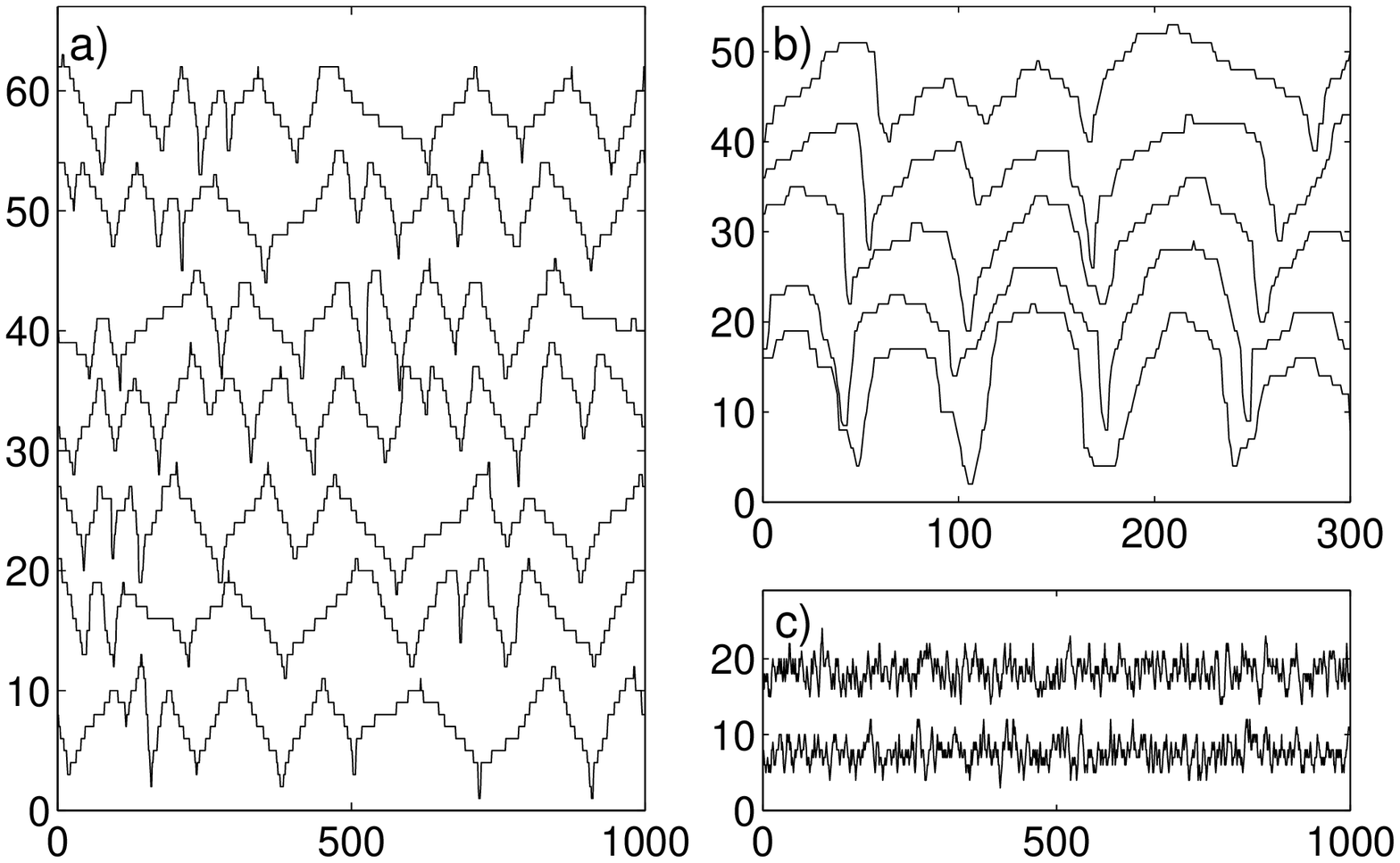,width=8.2cm}
\bigskip
\noindent {\bf Fig. 2}: Snapshots of typical step edge profiles with step orientations in the
(a) and (b) $\left[ 110\right]$
and (c) $\left[ 100\right]$ directions at $T=240$ K with
$F=6\times 10^{-3}$ ML/s for coverages $\theta=$1.0, 5.0, and 4.0,
respectively. In the latter case KESE vanishes and the instability
is strongly suppressed. The lateral and vertical scales are given in units
of the lattice constant.
\bigskip
\bigskip

\noindent $a=0.255$ nm, corresponding to a typical MBE regime \cite{Kru97}.
System sizes considered consisted of eight terraces, each of width $L_x=8.5$
and length up to $L_y=2000$. With fully periodic boundary conditions there was
a dependence of the selected wavelength on $L_y$. With open boundaries this
dependence disappears, and in the ledge direction there was no difference in
the results between $L_y=500$ and $L_y=1000$. However, periodic boundary
conditions were necessary to study the temperature dependence of the
structures, in which case we explicitly checked that this feature did not
depend on $L_y$.

The choice of the initial conditions warrants some discussion. In principle,
1D step edges are thermally rough in equilibrium. Thus, we chose to start
from initially rough step edges. We also performed some test runs starting
from ideal, smooth ledges. In these cases development of the instability was
much slower than when the steps were rough. We also discovered slight
dependence of the characteristic wavelength of the instability on the
initial condition. For both cases, however, we verified that the same
temperature dependence was obtained.

We first show results for the step orientation $\left[110\right]$ where KESE
is relatively strong. In Figs. 2 (a) and (b) there are step profiles after $%
\theta=1.0$ and 5.0 monolayers, respectively, at $T=240$ K with $F=6\times
10^{-3}$ ML/s. The development of an instability with a relatively
well-defined wavelength for all ledges is apparent, and the steps seem to
``phase-lock'' at the largest coverages studied. To quantify these results,
we consider the lateral height correlation function for step edge profiles $%
\zeta (x,\theta) \equiv h(x,\theta) - \overline{h}(\theta)$, where $h$ is
the step height measured from $y=0$, and $\overline{h}$ is its spatial
average: 
\begin{equation}
C(x,\theta)=\left\langle \zeta (x^{\prime},\theta)\zeta
(x+x^{\prime},\theta)\right\rangle.
\end{equation}
\psfig{file=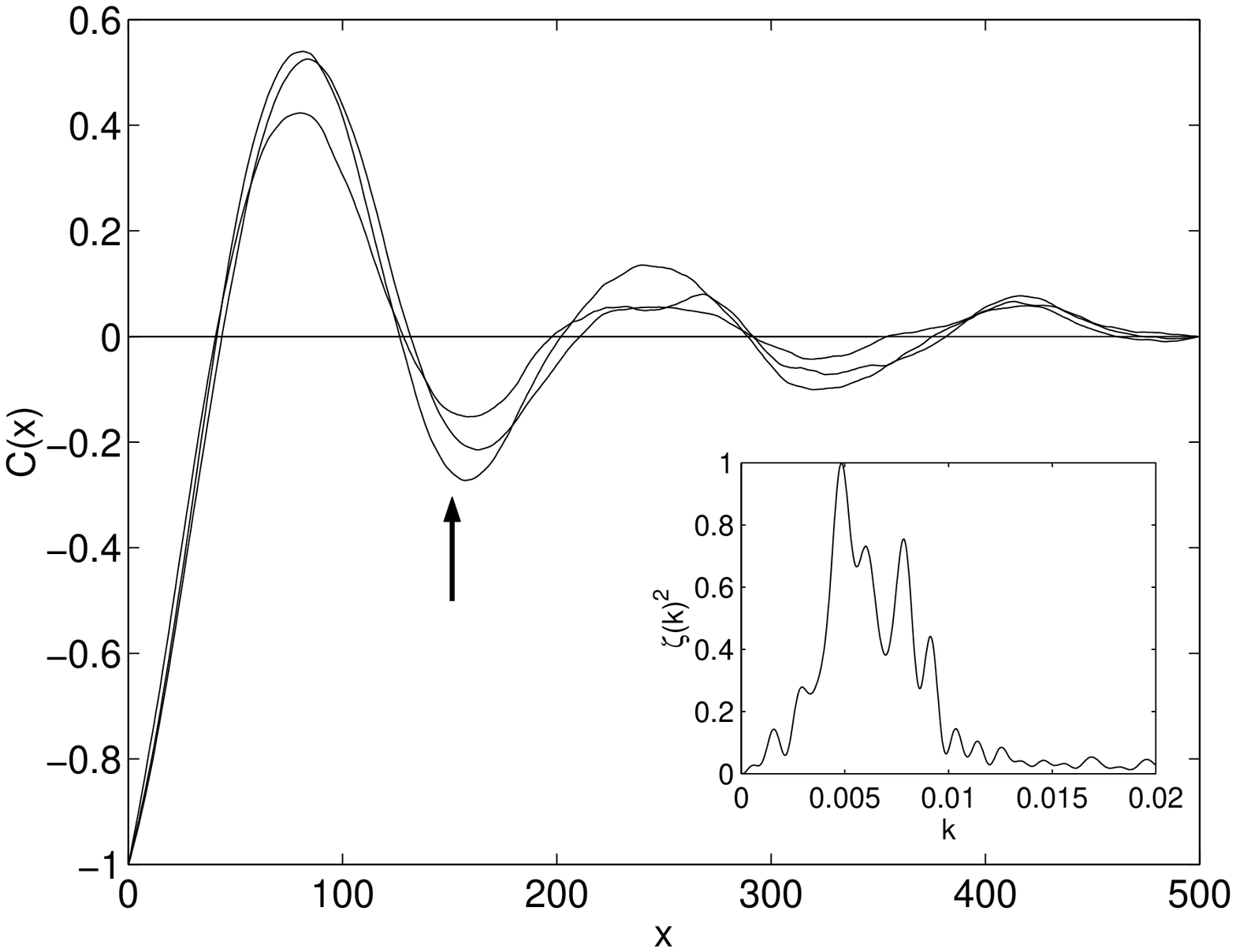,width=8.2cm}

\bigskip
\noindent {\bf Fig 3}:
The correlation function $C(x,\theta)$ for step edge fluctuations
along the $\left[ 110\right]$ ledge, for coverages
$\theta =$0.4, 1.0 and 2.0 at $T=270$ K. The corresponding Fourier 
components of $\zeta$ are shown in the inset. 
An estimate for the wavelength obtained from the average distances between
fingers is denoted by an arrow.
\bigskip

\noindent Here, the brackets denote an average over the (deposition) noise and
over all ledges in the system. The correlation function $C(x,\theta)$ for the 
$\left[ 110\right]$ step edges at $T=270$ K with $F=6\times 10^{-3}$ ML/s is
shown in Fig. 3 for $\theta =0.4,1.0$, and 2.0, and its Fourier components
are shown in the inset. The first minimum of $C(x,\theta)$ gives a measure
of the characteristic length scale $\lambda \approx 37$ nm in the system,
and as can be seen from Fig. 3 it is almost independent of $\theta$ even for
coverages below the apparent phase-locking. The same information is
contained in the Fourier components of $\xi$. The arrow in Fig. 3 shows an
estimate of the average distance between the growing fingers, as obtained
directly from the ledge configurations. We note that in experiments the
stabilization and phase correlation of the perturbations seems to improve
between 5 ML to 50 ML \cite{Mar99}. Due to computational restrictions we
have not been able to probe this regime, however.

In Fig. 2 (c) we show some step profiles for the open $\left[100\right]$
steps after $\theta=4.0$ monolayers. The difference with respect to the
close-packed steps is striking. Within the accuracy of the data, we find no
evidence of unstable growth but the ledge fluctuations appear completely
random. This is strong evidence indicating that in the present case, the
patterns on the $\left[110\right]$ ledges are due to KESE and not by BZI. In
the absence of KESE, enhanced edge diffusion along the sides of the steps
here strongly suppresses fluctuations of the $\left[100\right]$ ledges (cf.
Fig. 1 (b)).

Let us next consider the origin of the wavelength ($\lambda \approx 30$ nm
at $T=240$ K with $F=6\times 10^{-3}$ML/s) observed for the $\left[110\right]
$ ledges. It is of the same order of magnitude as the wavelength obtained in
the experiments \cite{Mar99}.
\psfig{file=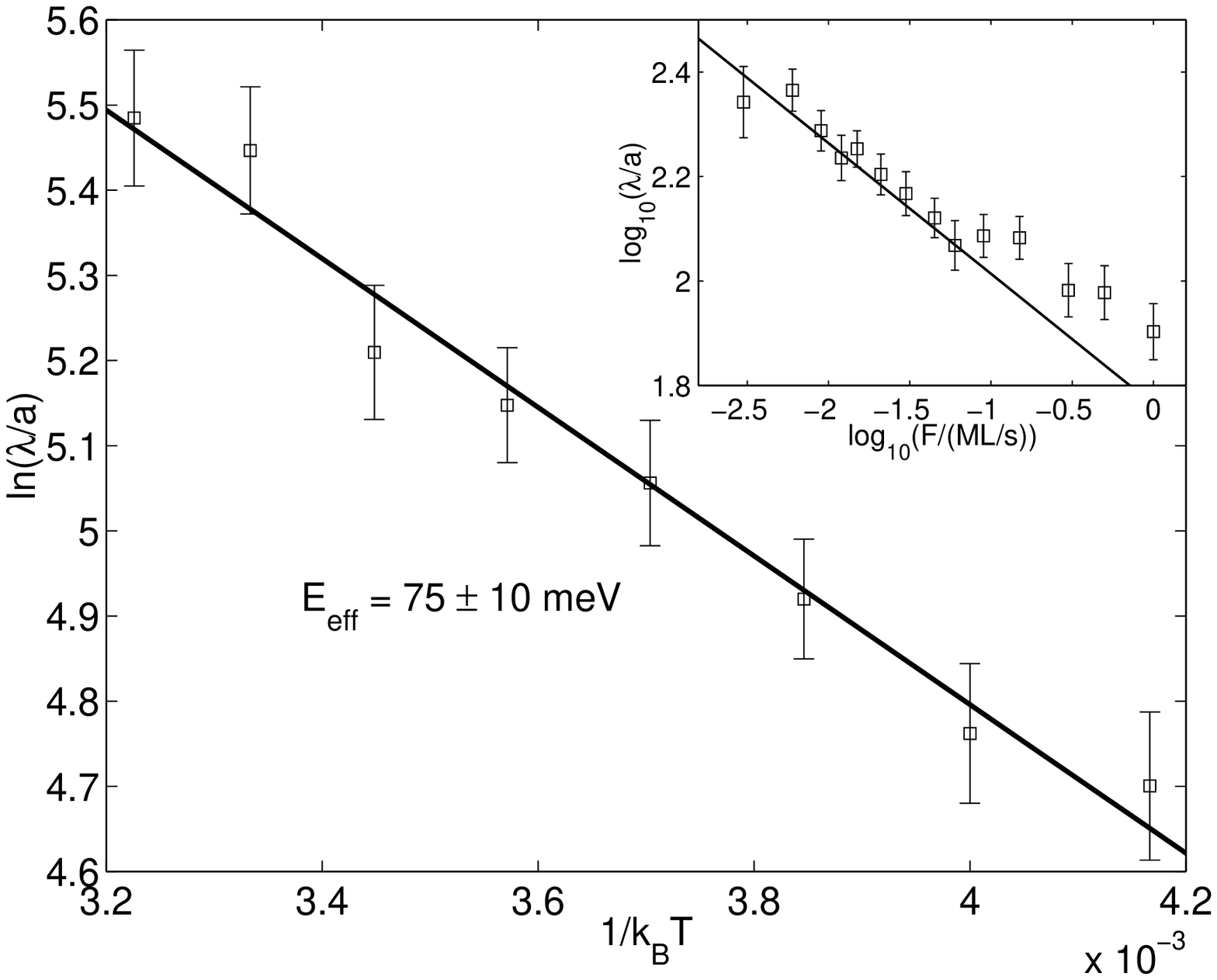,width=8.2cm}

\bigskip
\noindent {\bf Fig. 4}:
Temperature dependence of the the wavelength $\lambda$ obtained from
the distance between the fingers. It has been determined after deposition of
2.0 ML with $F=6\times 10^{-3}$ ML/s at $T=240 - 310$ K.
The least squares fit is shown by a continuous line with a slope
denoted in the figure.
In the inset the flux dependence of $\lambda$ is shown at $T=300$ K
with fluxes $F=3\times 10^{-3} - 1.0$ ML/s after deposition of 2.0 ML.
The straight line is the theoretical prediction of Eq. (3).
For the highest flux values, there is indication of terrace nucleation.
\bigskip

\noindent According to continuum theory for KESE \cite {Pie99},
there are two important length scales controlling step flow growth
with KESE. They are denoted by $\ell_{c}$ for dimer nucleation on a straight
step, and the the kink Schwoebel length $\ell_{s}= \exp [(E_{s}-E_{d})/k_BT]
-1$ \cite{Pie99}, which is related to the energy barriers $E_{s}$ and $E_{d}$
for jumps around a kink site and along a straight edge, respectively. When $%
\ell_{s}\ll \ell_{c}$, the nucleation length can be estimated from the model
of Politi and Villain \cite{Pol96} (see also Ref. \cite{Pie99}). However,
for the $\left[110\right]$ ledges, $\ell_s \simeq 10^4$ and $\ell_c \simeq
10^2$ around room temperature, which indicates that we must consider the
opposite case of a strong KESE.

In the case of a strong KESE, a more appropriate estimate for the nucleation
length $\ell _{c}$ is obtained by considering the probability for nucleation
of dimers at straight step edges. In the limit of infinite KESE it has been
shown that $\ell _{c}\sim (D_{s}/F_{s})^{1/4}$ \cite{Kru97b}. Here $D_{s}$ $%
\propto \exp [-E_{d}/k_{B}T]$ is the diffusion coefficient along the
straight ledge and $F_{s}=FL_{x}$ is the flux at the ledge. In the case of
finite but large $\ell _{s}$, the corresponding 1D diffusion equation with
appropriate boundary conditions has been solved by Politi \cite{Pol97}. From
his solution, assuming that $\ell _{c}$ is controlled by dimer nucleation at
the bottom of the growing step structure gives $\ell
_{c}=(12D_{s}/F_{s})^{1/4}$. Since the width of the 1D terrace fluctuates
between zero and $\ell _{c}$ \cite{Pol96}, we obtain an estimate for the
wavelength as 
\begin{equation}
\lambda \approx \frac{1}{2}\left( \frac{12D_{s}}{F_{s}}\right) ^{1/4}.
\end{equation}
The scaling exponent $1/4$ is a direct consequence of dimer
nucleation at the ledge and is always expected  for strong KESE within
the present model. The relation in Eq. (3) predicts that the
temperature dependence of $\lambda $ is controlled by an effective barrier $%
E_{{\rm eff}}=E_{d}/4$. In the present case this becomes about 65 meV. The
estimate for $E_{{\rm eff}}$ in the case of BZI gives a value nearly an
order of magnitude larger \cite{Mar99}. Eq. (3) also predicts that $\lambda
\sim F_{s}^{-\alpha }$ with $\alpha =1/4$, in contrast to the BZI case where 
$\alpha =1/2$ \cite{Bal90}.

With the present energetics, Eq. (3) gives $\lambda \approx 40$ nm at $T=270$
K with $F=6\times 10^{-3}$ ML/s which is in very good agreement with the
simulation result $\lambda \approx 37$ nm. This is in contrast to BZI which
yields $\lambda \approx 1$ nm \cite{Mar99}. We have furthermore tested the
predictions of Eq. (3) by estimating $\lambda$ at different temperatures and
also by varying the flux. In Fig. 4 we show the temperature dependence of $%
\lambda$ for various temperatures in an Arrhenius plot. A straight line fit
to the data gives $E_{{\rm eff}}= 75 \pm 10$ meV, in good agreement with the
prediction based on Eq. (3). This estimate is also in good agreement with
the experimental result by Maroutian {\it et al.} \cite{Mar99}. In the inset
of Fig. 4 we also show the flux dependence of $\lambda$ and compare with
that obtained form Eq. (3) with no fitting parameters. We find excellent
agreement between theory and simulations, except for the largest fluxes
where terrace nucleation becomes apparent. Fitting to flux values up to $%
F=9\times 10^{-2}$ ML/s gives $\alpha=0.23 \pm 0.03$.

Finally, we discuss the roughness of the step in terms of the width of the
step edge fluctuations $w(\theta )=\sqrt{\left\langle \zeta (\theta
)^{2}\right\rangle }$. We find that $w(\theta )$ increases with coverage and
does not show signs of saturation up to about 5 ML. Up to this region
the width follows a power law behavior $w(\theta )\sim \theta
^{\beta }$, with $\beta \approx 0.3$ as seen in case of isolated
step \cite{Sal93}. The solid-on-solid model gives for strong KESE an
exponent $\beta \approx 0.57$ \cite{Pie99}, and $\beta =1$
has been obtained for a one-sided step growth model with infinite kink
barrier \cite{Hei98}. This indicates that $\beta $ depends on the details of
the model for unstable growth.

To summarize, the general situation regarding 1D ledge instabilities under
growth is a complicated one \cite{Pol00} (see also \cite{Sal93,Hei98}). The
simulations presented here give support to the view that on vicinal Cu(1,1,$m
$) surfaces the observed instability is due to KESE. This means that the
competing BZI is of no importance neither in length and time scales studied
here nor in those accessed in the experiments. In our simulations we have
been able to demonstrate the onset of the wavelength selection, in
qualitative agreement with continuum theoretical predictions. Moreover, it
was possible to connect the observed temperature dependence of wavelength to
the underlying energetics. This is in qualitative agreement with
experimental observations and theoretical predictions for strong KESE, and
clearly incompatible with BZI. Simulations also confirm the crucial role of
the dimer nucleation length in determining the length scale of the patterns.
Although we cannot conclusively demonstrate the phase-locking of the step
fluctuations, there is evidence that this indeed occurs at higher coverages.

Acknowledgements: We wish to thank T. Maroutian for providing us with
unpublished data, and M. Rost and J. Kallunki for useful comments. This work
has been supported in part by the Academy of Finland through its Center of
Excellence program.

\end{document}